# Codeless Screen-Oriented Programming for Enterprise Mobile Applications

Aharon Abadi, Yael Dubinsky, Andrei Kirshin, Yossi Mesika, Idan Ben-Harrush

IBM Research – Haifa
{aharona,dubinsky,kirshin,mesika,idanb}@il.ibm.com

## ABSTRACT
Designing the user interface (UI) of mobile applications in the enterprise is performed in many cases by the applications' users who represent the business needs. This is based on existing developed enterprise services that can be accessed by these applications. Design the UI of the mobile applications require programming environments that are as much as possible codeless and easy to use. In this paper, we present NitroGen which is a cloud-based platform-independent tool that provides with no installation, a consumable integrated set of capabilities to construct mobile solutions aiming at reducing development and maintenance costs. We further illustrate how to use NitroGen showing its visual touch-based mostly codeless way of programming. NitroGen can easily connect to back-end services thus enable fast and facile development in enterprises.

## Categories and Subject Descriptors
D.2.2 [**Design Tools and Techniques**]: *User interfaces;* D.2.6 [**Programming Environments**]: *Graphical environments*

## 1. INTRODUCTION
Contemporary approaches in enterprises include the definition and design of mobile applications to reflect the business goals. In many cases, applications' users design the user interface of these applications while connecting to existing services e.g., for authentication and for organizational data.

Developing an application involves obsessing over vast amounts of technical details due to today's fragmented and proprietary mobile market [1, 2]. The existing state-of-the-art requires a deep understanding of technologies and internals, and the mobile aspect is adding a new set of challenges such as security, connectivity, and state synchronization. There are known existing development environments for mobile, among them Xcode [3], ADT [4], Titanium [5], PhoneGap [6]. These environments are for developers and require deep knowledge of the targeted technology. Further, these are not cloud-based and code-less environment. There are also touch based environments such as TouchDevelop [12]. The most important features of NitroGen comparing to other tools are that NitroGen is zero-installation cloud based tool and that it provides easy access to back-end services. We suggest that there is a need to define new models that focus on the goal of the application itself, provide interface for non-developers and enable facile connectivity with the enterprise services.

Based on previous work [7,8,9], we present the NitroGen tool. NitroGen is a visual, mostly codeless (drag and drop), cloud based environment to construct mobile applications. The tool uses enterprise managed interfaces and provides a consumable, high-level integrated approach to building mobile applications that are not specific to any single platform or device. NitroGen allows solutions to be quickly constructed from customizable templates, instead of developing them from scratch. This solution accelerates mobile application construction, lowers the costs, and simplifies the development process by eliminating the need for heroic efforts or deep technical skills.

In Section 2, we present NitroGen architecture. In Section 3, we describe its application abstraction and in Section 4 the service integration. In Section 5, we illustrate using NitroGen to construct a mobile application including connecting it to a service, and deploy it for both Android and iOS. In Section 6, we conclude.

## 2. NITROGEN ARCHITECTURE
IBM Mobile Foundation [10] delivers a range of application development, connectivity and management capabilities that support a wide variety of mobile devices and mobile app types. However, a great deal of mobile application development skills is still required to construct a mobile application. These skills include, but not limited to: programming languages (Objective C, Java, C#, HTML, JavaScript, CSS), development environments (XCode, Eclipse, ADT, Visual Studio), platform specific APIs, frameworks (Dojo, jQuery, Backbone.js), communication with back-end, mobile specific user experience. NitroGen aims to eliminate technical skills that are required to develop a mobile application and extends the IBM Mobile Foundation capabilities. It particularly is about making it easier to create line of business applications that use and create data within the enterprise. The target users of these mobile applications are themselves employees of the enterprise. These applications don't have the branding and experience requirements of customer facing applications. However, these days with more and better looking mobile applications around, the bar of what is considered a good mobile has significantly risen. NitroGen is all about exceeding the height of the bar quickly and easily.

## 2.1 Design time Components

IBM® WebSphere® Cast Iron® is a graphical tool that enables users to integrate cloud and on-premise applications. The results are exposed to mobile applications in a form of services, as illustrated in Figure 1.

Our contribution is based on IBM® Forms Experience Builder. With the help of this tool users of all skill levels can rapidly build and deploy data intensive web applications. NitroGen extends the tool for designing mobile applications. Services defined in Cast Iron® are dynamically discovered and bound to the application data fields. NitroGen allows the user to quickly preview the application in browser exactly as it will look like on a real device (various platforms are available). Then NitroGen generates the application code and publishes it to the IBM® Worklight [11] for validation and testing on actual mobile devices.

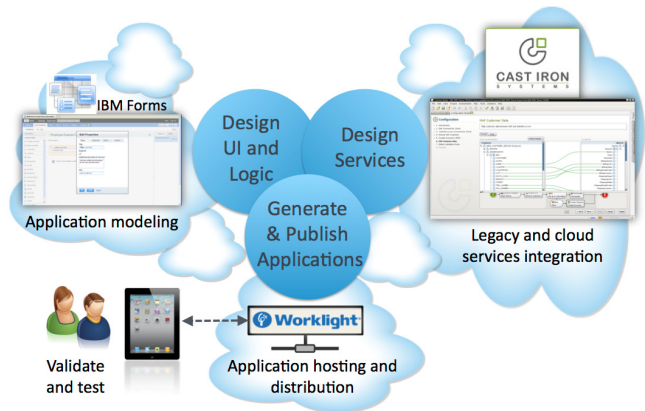

Figure 1: Design time Components

## 2.2 Runtime Components

Tested applications are hosted and distributed using IBM® Worklight. Applications securely communicate with Cast Iron® via Worklight server. This is illustrated in Figure 2. Cast Iron® in its turn brings together the data from various sources.

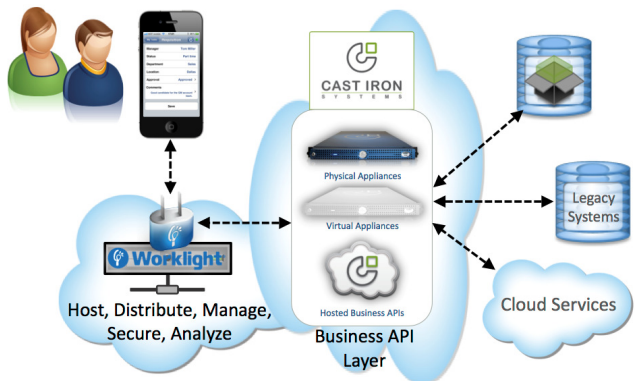

Figure 2: Runtime components

## 3. Application Abstraction

To deal with the complexity associated with mobile application development, we have defined an abstract application model that represents a mobile application. The user designs the application with the help of a simple and intuitive web based IDE never seeing any of the underlying models. Our runtime then interprets the application model providing the best user experience on different mobile platforms and form factors. While the created application is a true Model, View, Controller (MVC) type of application, the user of the IDE is unaware of this. He or she simply works on the User Interface (UI) and in a simple way connects the UI to backend data. The user is unaware of how the application connects to the backend data as the IDE uses meta-data about the various back end data sources to generate the needed Worklight artifacts for this.

An *application* is a set of *forms* (screens), each having a sequence of *fields* that show application data. Fields are of different *kinds*, such as text, telephone, date, and image. Fields can be either *read-only* to present data or *editable* to collect data. *Table* is a special kind of field for showing repetitive data. *Buttons* and *table rows* are used for *navigation* between forms. Under the covers these UI artifacts are bound to a data model that is extracted from the UI designed by the user.

*Services* transfer data between the application and back-end. The *pre-population service* is automatically invoked when a form opens to fetch the needed data. The *save service* is used to save form's data and its existence causes the runtime to automatically add a save button to the UI. The NitroGen IDE's server has an extendable method for "discovering" data sources available to the user when creating his/her application. This process provides information such as connectivity and data fields available enabling an easy to use User Experience for connecting to the needed back-end data sources.

When navigating between forms, data can be passed between the source form and the target form. *Global variables* are available in all forms and are used to pass data from one form to any other in subsequent navigations.

## 4. Back-end Service Integration

Although in some cases the mobile applications are standalone, in most of the cases interactions with backend services is essential. At runtime the mobile applications should integrate with backend service(s) to create, retrieve, modify and delete data. However, backend services are from a variety of domains and implements with various technologies introducing an implementation challenge to the application developer. The integration with the backend services in NitroGen is relevant in two stages of the application lifetime: design and runtime.

At design time the application developer is binding the services to the application. For this purpose the NitroGen platform should be able to provide a set of services that can be used by the application. However, the topology of backend systems may comprise more than one system that provides services each with potentially different technology. A discovery component is required to be able to communicate with each of the systems, using its own protocol, to fetch a list of services definitions. The service definitions should be a structured document that describes the available operations, input and outputs message formats as well as additional required information for the service invocation. The Cast Iron, for instance, is a backend system that provides SOAP based Application Program Interface (API) to query for available services and description of each service in a Web Service Definition Language (WSDL) format (see Figure 2). The result of the discovery component is a set of structured service description documents that contains the relevant information needed for the service binding. An instance of a description document consists of the service invocation protocol, system identifiers, input and output parameters (names and types), name, description and so forth. NitroGen platform use these service descriptors at design time to create the mapping dialog where developer is able to select the service and map fields to and from the service.

The same service descriptor is being used for generating the required adapters for the invocation of the service at runtime. Each adapter is configured to communicate with a specific backend system and acts as the mediator between the mobile application and the backend service. The main advantage of using an adapter is to take out the complexity of the integration from the client side to the server side. The use of an adapter is also crucial for overcoming security issues where the mobile application is not allowed to directly communicate with a backend service. In addition, the adapter transforms between the mobile application messages format and the backend service format.

## 5. ILLUSTRATING NITROGEN

In this part, we illustrate how NitroGen is step-by-step used to construct a mobile application including connecting it to a service, and deploy it for both Android and iOS.

The scenario deals with an enterprise that wishes to equip their aircrafts' technicians with a mobile application. The application enables the technicians to view their schedule and the details of a specific support ticket as well as filling in the details of their work for each specific ticket.

**Create a new application**

Nitrogen user defines the properties of a new application and creates a new one (as shown in Figure 3).

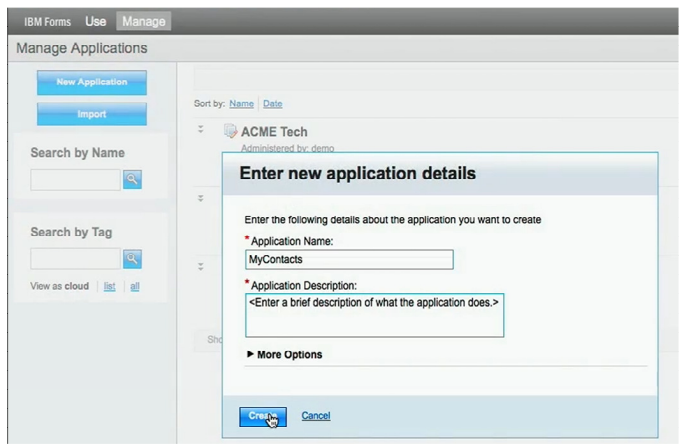

Figure 3: Create a new application

**Use the palette to develop the screens**

The user can use a palette with a variety of field types such as single line entry, date and text, to design the application screens. The palette contains also an address field that is connected to the location service, a phone number field that is connected to the phone itself, and a photo filed that is connected to the camera.

Figure 4 shows the palette interface. Figure 4 (a) shows the application layout at the upper right frame of the palette. We can observe five screens: Schedule, Customer, Ticket, Ticket History, and Summary, each of them has already one page. In Figure 4 (b), we can see that Page 1 of the Summary screen has few fields like Date, Status and so on.

Adding a new screen is done by to clicking the green plus sign as shown in Figure 5. This enables the user to select the desired screen layout.

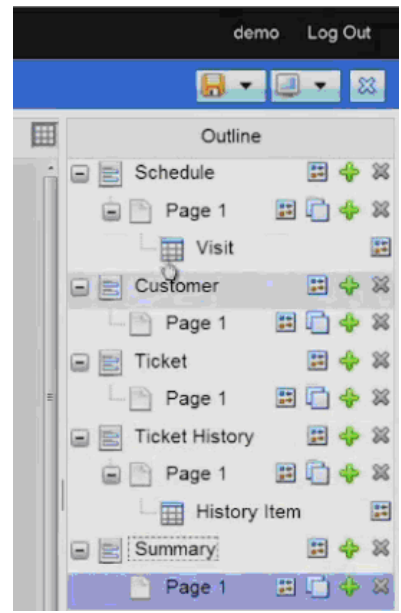
(a)

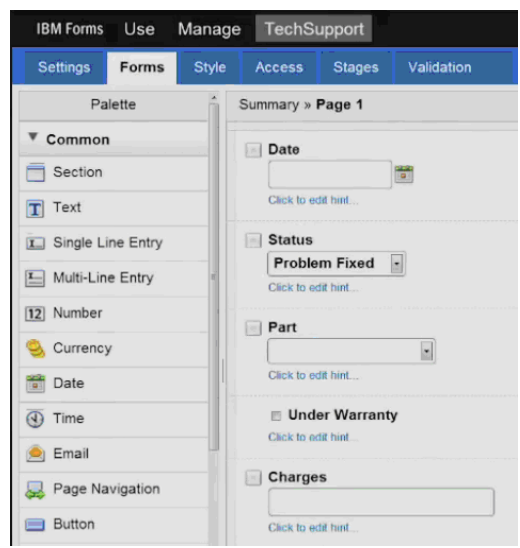
(b)

Figure 4: NitroGen palette

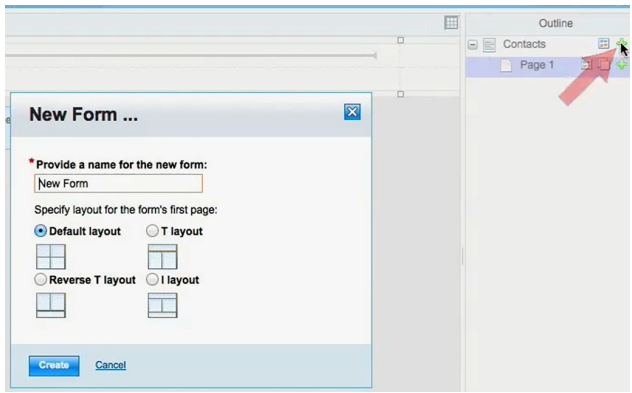

Figure 5: New screen

Adding a new field is done by dragging and dropping the field into the current screen. For example, in Figure 6, the user adds a table to the screen by dragging the table from the palette to the screen. At any time the user can rename an element by clicking on its name and edit its properties as shown in Figure 7.

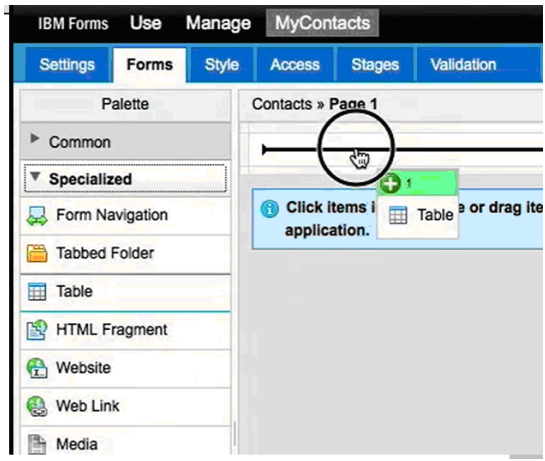

Figure 6: Adding a table field

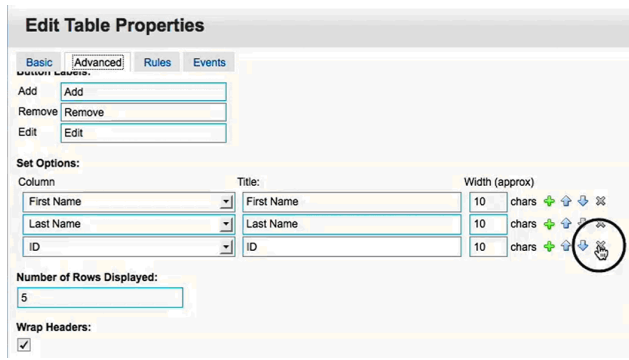

Figure 7: Table properties

Fields can be hidden by clicking on the 'x' sign at the right of the field definition as we can see at figure 7 still the data is a part of the application. The user does it for two reasons: (1) it is not part of the view; it is only for internal use of the application, (2) the mobile screen cannot show more than two columns.

**Connect to the enterprise services**

The user wishes now to connect to the enterprise services in order to populate screen fields with back-end data. As shown in Figure 8, by clicking on the screen property the user can open it.

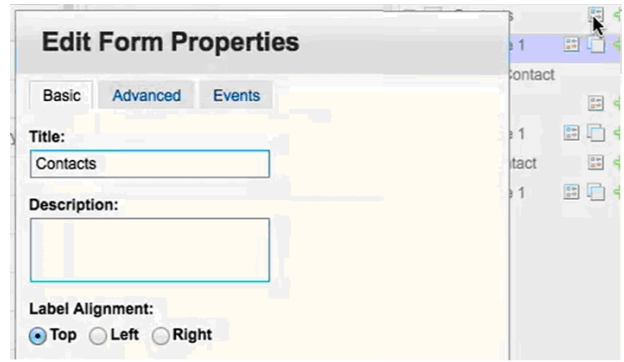

Figure 8: Edit screen properties

Moving to the 'Advanced' tab as shown in Figure 9, the user defines the service to be associated to this screen and selects the required service from the catalogue (Figure 10).

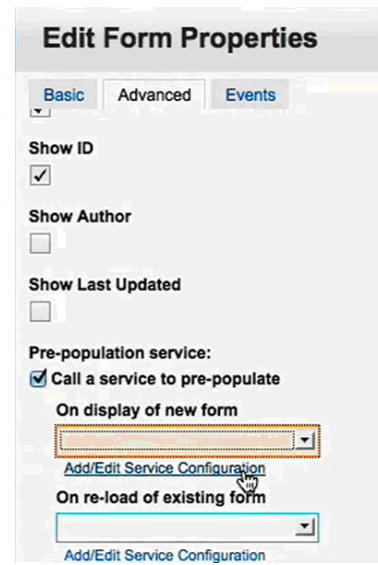

Figure 9: Connect to a service

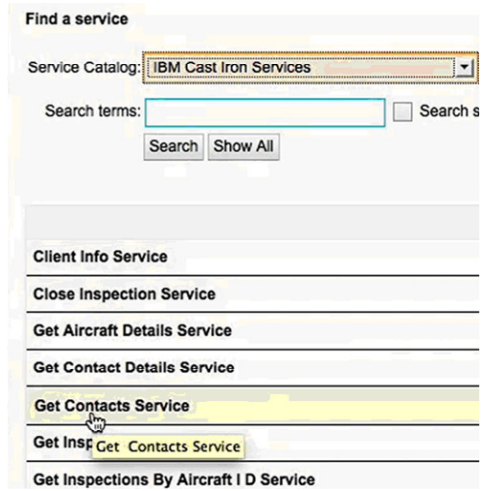

Figure 10: available services

The user maps inputs and outputs of the service to the screen fields. For example, in Figure 11, the 'Contact id' field of the service is mapped to the ID field in the table.

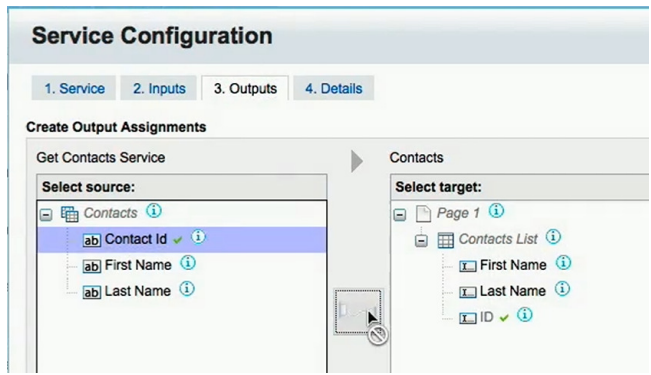

Figure 11: Data mapping

**Navigation between screens**

The user can easily define the navigation between screens. For example, in Figure 12, we see table navigation which is part of the table properties. Selecting this option, the user defines the target screen as shown in Table 13.

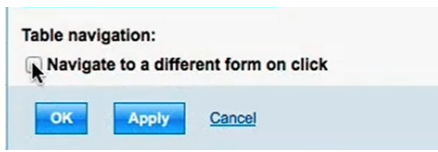

Figure 12: Table navigation

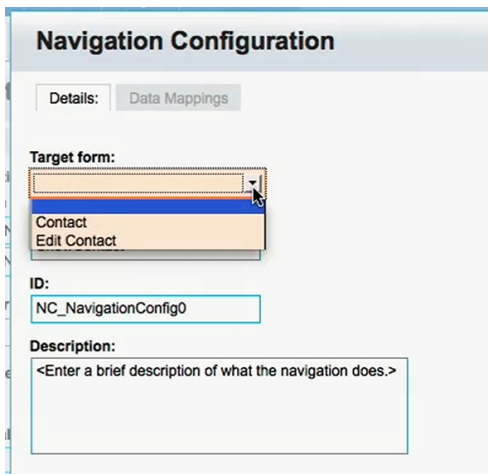

Figure 13: Navigation target screen

When navigating from one screen to another, the user can map the data fields of one screen to the target screen fields. For example, in Figure 14, the field 'Last Name' of one screen is mapped to the field 'Last Name' of the target screen.

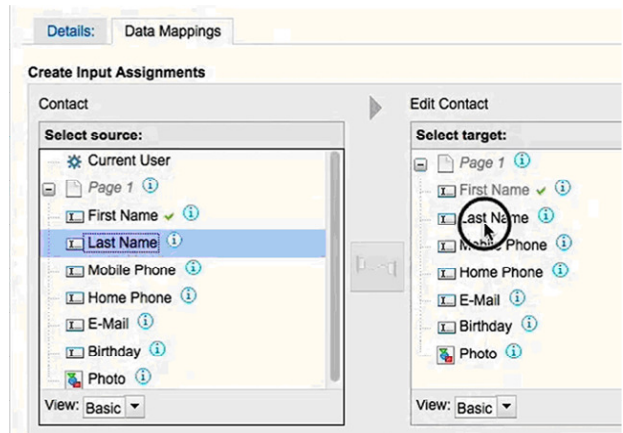

Figure 14: Navigation data mapping

**Deploy the application**

Finally, the user activates NitroGen application management in order to deploy the application. Clicking the 'deploy' option as shown in Figure 15, enables the user to define the deployment settings.

Figure 16(a) shows two 'TechSupport' applications that were generated, one for iOS and one for Android. These applications are available in the catalogue as shown in Figure 16(b) and are ready for installation and use on both platforms.

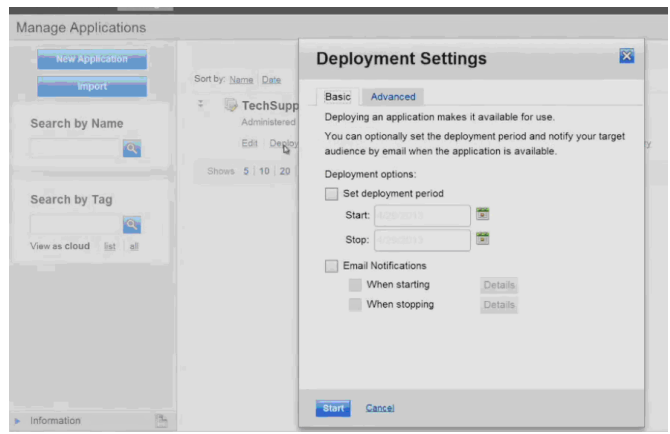

Figure 15: Deployment settings

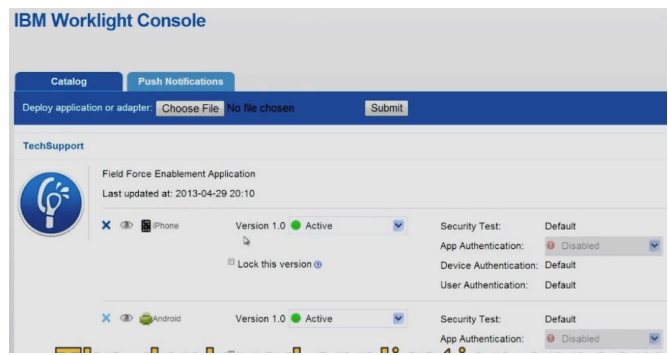

(a)

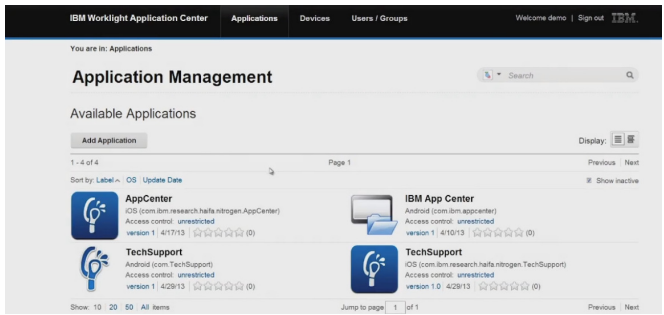
(b)

Figure 16: Deploy and archive for two platforms

## 6. SUMMARY AND FUTURE WORK
In this paper, we introduce NitroGen for constructing mobile applications that are not specific to any single platform or device and for supporting strong data transfer between the application and back-end services. Further, we illustrate using NitroGen for step-by-step development of a mobile application to be automatically deployed on both iOS and Android.

In preliminary evaluation, we found that NitroGen is easy to learn and use. Further, we note that most of NitroGen usage is based on touch mechanisms and therefore suits the approach that is used for mobile devices.

As future work we intend to evaluate NitroGen with both developers and non-developers in enterprise settings, and improve NitroGen to fit agile practices of testing, refactoring, and emerging design.

## 7. ACKNOWLEDGEMENTS
We thank Samuel Kallner for significant comments.

## 8. REFERENCES

[1] Gavalas,D.,Economou,D.,Development platforms for mobile applications: status and trends. IEEE Software 28(1),2011.
[2] Holzer A., Ondrus J., Trends in Mobile Application Development, Social Informatics and Telecommunications Engineering, Springer Berlin, pp. 55-64, 2009.
[3] Xcode, https://developer.apple.com/xcode.
[4] ADT, http://developer.android.com/-tools/sdk/eclipse-adt.html.
[5] Titanium, http://www.appcelerator.com/platform-/titanium-platform.
[6] PhoneGap, http://phonegap.com.
[7] Abadi A., Dubinsky Y., Kirshin A., Mesika Y., Ben-Harrush I., Hadad U., NitroGen: Rapid Development of Mobile Applications, SPLASH'13 Demonstrations, October 26-31, Indianapolis.
[8] Shachor, G., Rubin, Y., Guy, N., Dubinsky, Y., Barnea, M., Kallner, S., Landau, A. What You See And Do Is What You Get: A Human-Centric Design Approach to Human-Centric Process, Business Process Design (BPD), in BPM, 2010.
[9] Rubin Y., Kallner S, Guy N., Shachor G., Restraining Technical Debt when Developing Large-Scale Ajax Applications, WEB 2013, Seville, Spain.
[10] IBM Mobile Foundation [online] http://www-01.ibm.com/software/mobile-solutions/
[11] Worklight [online] http://www-03.ibm.com/software/products/us/en/worklight/
[12] Nikolai Tillmann, Michal Moskal, Jonathan de Halleux, Manuel Fähndrich, Sebastian Burckhardt: TouchDevelop: app development on mobile devices. SIGSOFT FSE 2012: 39